\documentclass[12pt,preprint]{aastex}
\def\lap{\lower.5ex\hbox{$\; \buildrel < \over \sim \;$}}
\def\gap{\lower.5ex\hbox{$\; \buildrel > \over \sim \;$}}

\def\ergcm2s{${\rm erg\ cm^{-2}\ s^{-1}}$}

\def\ergscm2s{${\rm erg\ cm^{-2}\  s^{-1}}$}

\def\cm-2{${\rm cm^{-2}}$}
\def\ergs{${\rm erg\ s^{-1}}$}

\def\sax1808{SAX~J1808.4-3658}

\def\lax    {${_<\atop^{\sim}}$}
\def\gax    {${_>\atop^{\sim}}$}
\def\t3 {r2-67}

\def\mathfont#1{\ifmmode{#1}\else{$#1$}\fi} 
\def\lae{\mathrel{<\kern-1.0em\lower0.9ex\hbox{$\sim$}}}  
\def\gae{\mathrel{>\kern-1.0em\lower0.9ex\hbox{$\sim$}}}  
\def\kms{\ifmmode{{\rm km\ s}^{-1}}\else{${\rm km\ s}^{-1}$}\fi}

\def\msun{\ifmmode{\ {\rm M}_\odot}\else{$ {\rm M}_\odot$}\fi}  
\def\msunyr{\ifmmode{\msun \ {\rm yr}^{-1}}\else{$\msun \ {\rm yr}^{-1}$}\fi}

\def\ref#1{\noindent\hangindent=24.0pt\hangafter=1{#1}\par}
\def\la{\hbox{\rlap{$<$}\lower.5ex\hbox{$\sim$}\ }}
\def\ga{\hbox{\rlap{$>$}\lower.5ex\hbox{$\sim$}\ }}
\def\lap{\lower.5ex\hbox{$\; \buildrel < \over \sim \;$}}
\def\gap{\lower.5ex\hbox{$\; \buildrel > \over \sim \;$}}

\def\sgra   {Sgr~A$^*$}
\def\innerseven {M31GC~J004246+411737} 
\def\none		{CXOM31~J004244.3+411608}	
\def\sss	{CXOM31~J004244.3+411607}

\def\kms    {~km~s$^{-1}$}

\def\ergscm2s  {~erg~cm$^{-2}$~s$^{-1}$}
\def\cm2s   {~cm$^{-2}$~s$^{-1}$}
\def\cm2   {~cm$^{-2}$}
\def\cm3   {~cm$^{-3}$}

\def\chandra	{{\it Chandra}}

\received{2004 xx xx}
\begin{document}

\shortauthors{Garcia et al.}
\shorttitle{M31* with Chandra}
\slugcomment{submitted to {\em The Astrophysical Journal}}

\title{A Possible Detection of M31* with \chandra}

\author{Michael R. Garcia\altaffilmark{1}, 
Benjamin F. Williams\altaffilmark{1}, 
Feng Yuan\altaffilmark{2}, 
Albert K. H. Kong\altaffilmark{1}, 
F. A. Primini\altaffilmark{1}, 
P. Barmby \altaffilmark{1}, 
P. Kaaret \altaffilmark{3},
and Stephen S. Murray\altaffilmark{1}}
\altaffiltext{1}{Harvard-Smithsonian Center for Astrophysics, 60
Garden Street, Cambridge, MA 02138; williams@head-cfa.harvard.edu;
garcia@head-cfa.harvard.edu; akong@head-cfa.harvard.edu;
fap@head-cfa.harvard.edu; 
ssm@head-cfa.harvard.edu} 
\altaffiltext{2}{Department of Physics, Purdue University, 525
Northwestern Avenue, West Lafayette, IN 47907;
fyuan@physics.purdue.edu}
\altaffiltext{3}{Department of Physics and Astronomy, University of
Iowa, Iowa City, IA 52242-1479,  philip-kaaret@uiowa.edu}

\begin{abstract} 

Two independent sets of \chandra\ and HST images of the nuclear region
of M31 allow registration of X-ray and optical images to $\sim 0.1''$.
This registration shows that none of the bright ($\sim 10^{37}$\ergs)
X-ray sources near the nucleus is  coincident with the central
super-massive black hole, M31*.  A 50ks \chandra\ HRC image shows 
2.5$\sigma$ evidence for a faint ($3 \times 10^{35}$\ergs),
apparently resolved source which is consistent with the position of
the M31*.  The Bondi radius of M31* is $ 0.9''$, making it one of the
few super-massive black holes with a resolvable accretion flow.  This
large radius and the previous detections of diffuse, X-ray emitting
gas in the nuclear region make M31* one of the most secure cases for a
radiatively inefficient accretion flow and place some of the most
severe constraints on the radiative processes in such a flow.
 
\end{abstract}


\section{Introduction}

It is now accepted that a
supermassive black hole (SMBH) is present in essentially every galactic
core~\citep{smbh.everywhere}. However, most galactic nuclei 
appear inactive, raising  the 
question: ``Why are these SMBHs such embarrassingly faint X-ray
sources, as compared to the SMBHs previously known to exist in AGN?''
Explanations fall into two broad classes - either 
the accretion rate is extremely low, i.e., the SMBHs are  
'starved', or the accretion process is radiatively inefficient. 
A straightforward way to determine which explanation is correct is to 
resolve the accretion
flow and therefore securely determine the mass accretion rate.
We note that hybrid explanations are possible, ie, the flow could
start at the Bondi rate and then be slowed or stopped by winds,
convection, and/or magnetic fields~\citep{perna.b}. 

Because of the limited resolution of current X-ray telescopes the only
currently known examples of resolved accretion flows into SMBHs are
those in \sgra\ \citep{baganoff.03} and M87 \citep{dimatteo.m87}.
Assuming the accretion flow in \sgra\/ has the ``normal'' accretion
efficiency, its expected luminosity would be roughly seven orders of
magnitude greater than the observed value.  So we cannot escape the
conclusion that the accretion flow in Sgr A* is radiatively
inefficient \citep{yuan.sgra.03}.  M87 is under-luminous only by four
(not seven) orders of magnitude so must also have a radiatively
inefficient flow but the presence of a strong and resolved nuclear jet
complicates the picture. 

At a distance of 780~kpc~\citep{stanek.780kpc,macri.780kpc}, M31$^*$ is
the nearest analog to Sgr A*.  In addition to its   highly unusual double
nucleus, the center of M31 houses a $3\times 10^7$~\msun\/ black hole
\citep{kormendy.bender.99} spatially coincident with the point radio
source M31* \citep{crane.92}. The radio source is unresolved at the
$\sim 0.35'' \sim 1$pc level.  While M31* is 100x further away than
\sgra, it suffers much less reddening: $A_V \sim 1$
\citep{garcia.m31*.2000}, whereas $A_V \sim 30$ for \sgra.

The first \chandra\ observations of M31 led us to associate a
super-soft source near M31* with this SMBH \citep{garcia.m31*.2000}.
Subsequently we were able to register several HST WFPC2 images of the
nuclear region with our ACIS mosaic using two globular clusters,
reducing the error circles of the X-ray sources by a factor of $\sim
10$.  This showed that our initial association was incorrect, and that
none of the bright (${\rm L_X} \sim 10^{37}$\ergs) sources near the
center of the galaxy is spatially coincident with M31*
\citep{garcia.m31*not.2001}.  Recently we have obtained an HST ACS
image which fortuitously contains M31*.  Registration of this image
with a \chandra\ HRC image confirms our earlier results, and suggests
that a previously unresolved faint source may be co-incident with
M31*.  In this paper we present both datasets and discuss M31* in the
context of other low-luminosity SMBH.

\section{Observations}

The absolute astro-metric accuracy of both \chandra\ and HST images is
only $\sim 1''$, but by using common sources one can register the
images to $\sim 0.1''$.  As globular clusters are some of the most
common X-ray point sources found in M31 and are easily identifiable on
HST images, we searched for clusters near the nucleus which could
provide the desired registration.  We found two such clusters. 

Below we show that we were then able to accurately register two
completely independent datasets, one using the HST/WFPC2 and the
\chandra/ACIS-I, and the other using the HST/ACS and the \chandra/HRC.
Both confirm that none of the bright (${\rm L_X} \sim 10^{37}$\ergs)
sources near the nucleus are coincident with M31*.  The HRC image
shows marginal evidence for a weak, resolved X-ray source at the
position of M31*.  While the ACIS observations do not show a resolved
source, they are consistent with a source of the same flux at the M31*
position.

\subsection{HST WF/PC2 and Chandra ACIS Observations}

The \chandra\ observations of the nuclear region analyzed herein
consist of 7 separate ACIS observations taken during the first two
years of \chandra\ operations, which have been summed to generate an
image with an effective exposure time of 34.7~ks.  These observations
are described in detail by \citet{kong.2002}, who considered these 7
and one additional (OBSID 1583) observation.  In order to maintain the
highest possible spatial resolution in the summed image, the
individual images were first generated in $1/8''$ pixels, with the
standard $\pm 1/4''$ position randomization removed.  The images where
then stacked using the positions of the point X-ray sources as
registration marks.  The resulting image has a radially averaged FWHM
on axis of $\sim 0.6''$ as determined by IRAF\footnote{IRAF is distributed by the National Optical Astronomy
Observatory, which is operated by the Association of Universities for
Research in Astronomy, Inc., under cooperative agreement with the
National Science Foundation.} (imexam) profile fitting
on numerous sources.  We limited the image size to $2048^2$ pixels,
therefore covering $4.2' \times 4.2'$.  The central region of this
image is show in Figure~1.
Two of the three bright nuclear
sources are visible in this image, including the northernmost source
(N1 = \none / r1-10) and the central super-soft source (SSS = \sss /
r1-9, \citealp{kong.2002}).

An archival WFPC2 image which includes M31$^*$ also includes a
recently discovered globular cluster (\innerseven) identified by
\citet{barmby.huchra.2001}.  This archival image was taken with the
F300W filter (central wavelength 3000\AA, bandpass 300\AA) on 1995
June 19.  We obtained a WFPC2 image on 2000 Feb 02 in order to search
for a UV counterpart the the new {\it Chandra} X-ray transient CXOM31
J004242.1+411608 (\citealp{murray.m31trans, kong.2002}; also called CXO
J004241.0+411608 in \citealp{garcia.m31*.2000}).  This image was taken
with the F336W filter (central wavelength 3360\AA, bandpass 200\AA)
and also includes M31$^*$, as well as the cluster mita213 in the
catalog of \citet{magnier.phd}.  Both of these globular clusters are
also detected in the merged ACIS-I image, as sources r2-15 and r1-32
in \citet{kong.2002}.

These two HST images were taken at orientations that differ by nearly
180$^o$, allowing them to cover the maximal area of M31, while still
allowing partial overlap of the PC and WFC chips.  In these
overlapping regions we identified $>100$ individual point sources,
which we used to register the two images to a common frame.  We
arbitrarily choose the frame of the F300W image as determined by the
HST pipeline processing as our preferred reference frame.  The
registration of the two WFPC2 images was done with standard IRAF
routines (mkpattern, ccsetwsc, wregister, imcombine, geomap, geotran,
wsccopy; as described on the wregister help page) and has a rms error
of $0.05''$.  The mosaic of F300W and F336W images covers
approximately $3' \times 3'$.  The central region of the F300W PC chip
was centered on M31* and the central part of this PC image is show in
Figure~2.  The ACIS contours from Figure~1 are overlayed.

\subsubsection{Registering the WFPC2 and ACIS Images}

The errors in the registration of the \chandra\ image to the HST
frame are dominated by the positional uncertainties of the globular
clusters as measured in the \chandra\ image.  These errors are $r =
FWHM/(2.3*\sqrt{N})$ where N is the number of counts in each cluster.
From the images, we measure $r=0.06''$ for mita213 (with 59 counts)
and $r=0.07''$ for  \innerseven (with 39 counts).  We note that the
\chandra\ wavedetect tool gives similar errors.

Of course, the position of the globular clusters within the HST image
also has some error.  The $FWHM$ of the images of these clusters
themselves, as measured from the HST images is $0.50''$ for mita213
and $0.35''$ for \innerseven .  However, the much larger number of counts
in the HST images (\gax 10,000) indicates these position errors are
\lax~$0.01''$.
Summing the rms of the HST fit and the
errors for the two clusters used for registration in quadrature
indicates a 1$\sigma$ error for position determination of $0.082''$
which we conservatively round to $0.1''$.  We note that this
registration was done with a simple translation.  We did not find any
errors in roll or scale factor, which indicates they are smaller than
our computed registration error.

\subsection{HST ACS and \chandra\/ HRC Observations}

An HST ACS image of the M31 nucleus was obtained by us on 2004 Jan 23
as part of our \chandra\/ AO5 program searching for the optical
counterparts of black hole X-ray novae in M31.  The 2200~second image
was obtained with the F435W (=B band) filter in a standard 4 point
dither pattern and reduced with standard HST drizzle tools.  Figure~3
shows the region of this image surrounding M31$^*$, overlayed with
X-ray contours from the archival 47~ks HRC-I observation shown in
Figure~4 (OBSID 1912, obtained on 2001 Nov 1,
see~\citealp{kaaret.hrc}).  The contours in Figure~3 show the
super-soft source (SSS) on the bottom with source N1 above it.
Approximately $0.5''$ to the right a separate contour is seen
overlaying M31*.

Figure~4
shows the 47ks HRC image at its intrinsic $1/8''/$pixel
scale.  At a position consistent with M31* there appears to be a weak
source which is resolved from N1 and SSS.  There are 13 counts within
a 3 pixel diameter circle, which when assuming the typical M31 source
spectrum of ${\rm N_H}=7\times 10^{21}$cm$^{-2}$ and $\alpha = 1.7$
corresponds to a luminosity of ${\rm L_X} = 3 \times 10^{35}$\ergs\/.  The
background at this location is enhanced due to the scattering wings of
N1 and SSS and due to the unresolved diffuse emission in the core of
M31.  By summing the counts in 3 pixel diameter circles placed in a ring
equidistant from N1, we determined the background contribution from N1
and the diffuse emission to be $3.4 \pm 2.3$ counts at the location of
M31*.  In the same way we determined the contribution from SSS to be
$2\pm 1.9$ counts.  The 13 counts at the position of M31*  therefore
represents a $\sim 2.5\sigma$ detection.  While very suggestive, this clearly
needs confirmation.   Alternatively we may use the data to derive a
$3\sigma$ upper limit to the luminosity that is only slightly higher
at ${\rm L_X} = 3.6 \times 10^{35}$\ergs .

\subsubsection{Registering the ACS and HRC Images}

The ACS image described above contains both mita213 and M31$^*$, but
no other globular clusters.  We therefore used mita213 to register the
two images, and using the same type of error analysis as for the
ACIS-I/WFPC images we find a 1$\sigma$ position error of $0.11''$.  We
note that because we are limited to a single registration object in
this case we are not able to check for roll or plate scale errors.
However, the fact that these were measured to be negligible for the
ACIS-I/WFPC case indicates that they should be negligible here as
well.

\section{Discussion}

Having resolved M31$^*$ from the surrounding point and diffuse X-ray sources
we can now investigate the accretion properties of this nearby SMBH.  The 
presence of hot and truly diffuse emission component in the core
of M31 was first noted in {\it Einstein}\/ observations \citep{tf.m31}
and later confirmed with {\it ROSAT}\/ \citep{fap.rosat.m31}, {\it
XMM-Newton}\/ \citep{shirey.xmm}, and 
{\it Chandra}\/ \citep{anil.diffuse} observations.   Any of this 
gas within the Bondi radius of the SMBH will accrete and possibly
generate accretion luminosity. 
In order to compute the
Bondi accretion rate we use the X-ray observations to estimate the
 temperature and density of this gas.
  
The temperature in the inner $\sim 5'$ has been estimated as $\sim
0.35$~keV based on {\it XMM-Newton}\/ observations \citep{shirey.xmm}
and been found to  fall radially from 0.50~keV to 0.26~keV based on {\it
Chandra}\ ACIS-I observations \citep{anil.diffuse}.  Recently
\citet{taka.m31} has combined {\it XMM-Newton}\ and {\it Chandra}\
ACIS-S observations to reveal that there are three distinct soft
temperature components which have slightly different radial surface
brightness variations, therefore explaining the slight discrepancy
between the earlier two results.  The dominant component in terms of
emission measure and particle density has a temperature of $\sim
0.3$~keV.

The density of the diffuse gas near the SMBH can be estimated from the
emission measure and surface brightness profile fits.  \citet{anil.diffuse}
measured the profile within the central $1'$ and find that it appears
to flatten.  They estimate a gas density of $\rho \sim 0.1$~cm$^{-3}$. 
\citet{taka.m31} measure a separate surface brightness profile for
each of the three temperature components they find, in $1'$ bins.
Summing the flux from all three in the inner bin also indicates a
density of $\sim 0.1$~cm$^{-3}$ within $1'$ ($\sim 200$ pc) of
M31$^*$.   We note that this density is similar to that found in other
nearby galaxies \citep{dimatteo.m87, lowen.smbh}, but a factor of ten
below that found near \sgra \citep{baganoff.03}.

In computing the Bondi radius we follow the prescription of Baganoff
et al. (2003) exactly, thereby allowing direct comparisons. The radius
is $R_{\rm B} = 2 GM_{BH}/c_s^2$, where the sound speed $c_s = (\gamma
kT/\mu m_H)^{1/2}$.  Here the adiabatic index $\gamma = 5/3$ and the
mean atomic weight of the gas $\mu = 0.7$.  Given the density and
pressure in the previous paragraphs and using the dominant temperature
component of 0.3~keV,  we find for M31$^*$ 
$R_{\rm B} = 3.4$~pc, or $R_{\rm B}= 0.9''$ at the 780 kpc distance of M31.  

This can be compared with \sgra\/ with $R_{\rm B} = 0.072$~pc or
$=1.8''$ \citep{baganoff.03} and M87 with $R_{\rm B} = 1.7''$
\citep{dimatteo.m87}.  Loewenstein et al. (2001) have computed the
Bondi radii for a set of nearby elliptical galaxies in which the
temperature of the central diffuse gas can be measured, and find
$0.049'' < R_{\rm B} < 0.36''$ for this set.  Assuming similar
temperatures, we computed the apparent angular radii of the SMBH in
the definitive study of such objects in nearby galaxies (Magorrian et
al. 1998) and find in all cases $R_{\rm B}$\lax$0.4''$ ($R_{\rm B}
\sim 0.4''$ only for NGC 4486, NGC 4649, and NGC 4889; for the
majority of the objects $R_{\rm B} < 0.1''$).

While resolving the Bondi radius allows one to securely measure the
magnitude of the accretion flow, the next most fundamental number is
likely the ratio of the observed luminosity to the Bondi luminosity
${\rm L_{\rm B}} = 0.1 \dot{M_{\rm B}} c^2$.  The Bondi luminosity is
the maximum luminosity that one could expect, assuming that the
initially spherical flow eventually 
forms a classical thin accretion disk.  Following Baganoff,
$\dot{M_{\rm B}} = \pi(GM_{BH})^2 \rho c_s^{-3} = 3.4 \times 10^{20}
{\rm g~s^{-1}}$ where $\rho = n_e \mu m_H$.  For M31$^*$ we compute
${\rm L_{\rm B}} = 3 \times 10^{40}$~\ergs .  The possible detection
of M31$^*$ indicates ${\rm L_X} = 3 \times 10^{35}$~\ergs\/ and
therefore ${\rm L_X/L_{\rm B}} = 1 \times 10^{-5}$.

The Bondi radii and ${\rm L_X/L_{\rm B}}$ for nearby SMBH are plotted
in Figure~5.  This figure shows that M31$^*$ is one of only three
nearby SMBH with a Bondi radius larger than the {\it Chandra}\/
resolution.  These objects therefore allow the most secure
measurements of the accretion rate at the outer edge of the flow.
Knowing this, we may be able to determine if the flows contain winds
(ADIOS) convection (CDAFs), or if the emission is dominated by jet
processes (Falcke 2001) rather than thermal emission.  For example,
\citet{gpf.0.7} recently found that in quiescent stellar mass black
holes there is a correlation between the radio and X-ray luminosity
${\rm L_R \propto L_X^{0.7}}$.  The correlation also applies in AGN
\citep{merloni2003} but the normalization of the correlation depends
on the mass of the SMBH powering the AGN. Yuan and Cui (2005) model
these correlations in terms of  emission from a jet, predicting that
the correlation should steepen at the very low luminosity found here
for M31$^*$.  The observed X-ray and radio luminosity of M31$^*$
fall on this steeper correlation of ${\rm L_R \propto L_X^{1.23}}$
\citep{yuan.cui.2005}.  Interpreted in terms of this model, our
observations 
imply that both the X-ray and radio emission from M31$^*$ are
dominated by emission from an un-resolved, low bulk Lorentz factor
($<1.2$) outflow. 

While objects on the extreme right of Figure~5 provide the most secure
measurements of the accretion rate because ${\rm R_{\rm B}}$ is resolved,
those at the extreme lower end provide the most severe constraints on
the accretion processes because the flows are the most under-luminous.
We see that M31$^*$ is exceeded only by \sgra\/ in providing
simultaneously the most secure and most severe constraints.  
If our possible detection is incorrect, then the true luminosity
of M31$^*$ must be even lower.  As an example, we show in Figure~5 the
upper limit possible from a 200~ks Chandra HRC-I observation, which is
a factor of 10 below our possible detection. We plan such a deep
observation during Chandra~AO6.   Such a low ${\rm L_X/L_R}$
ratio  may be difficult to explain in terms of the jet model proposed
by \citet{yuan.cui.2005}. 

\section{Summary}

The observations we present herein provide evidence for a marginal
detection of X-ray emission from M31$^*$ at the level of ${\rm L_X} = 3
\times 10^{35}$\ergs, which is $10^{-5}$ times below the expected
Bondi accretion luminosity.  The Bondi accretion radius of the diffuse
X-ray plasma surrounding M31$^*$ is $R_{\rm B} = 0.9''$, larger than
the resolution of the X-ray observations.  The estimated Bondi
accretion rate is therefore very securely known.  These two points
taken together show that M31$^*$ is exceeded only by \sgra\/ in
providing the most secure and severe evidence for a radiatively
inefficient accretion flows in SMBHs.

This work was supported in part by NASA LTSA grant NAG5-10889 and
Contract NAS8-39073 to the Chandra X-ray Center. 

\vspace{-0.5cm}
\bibliography{apjmnemonic,references}
\bibliographystyle{apj}

\begin{figure}
\plotone{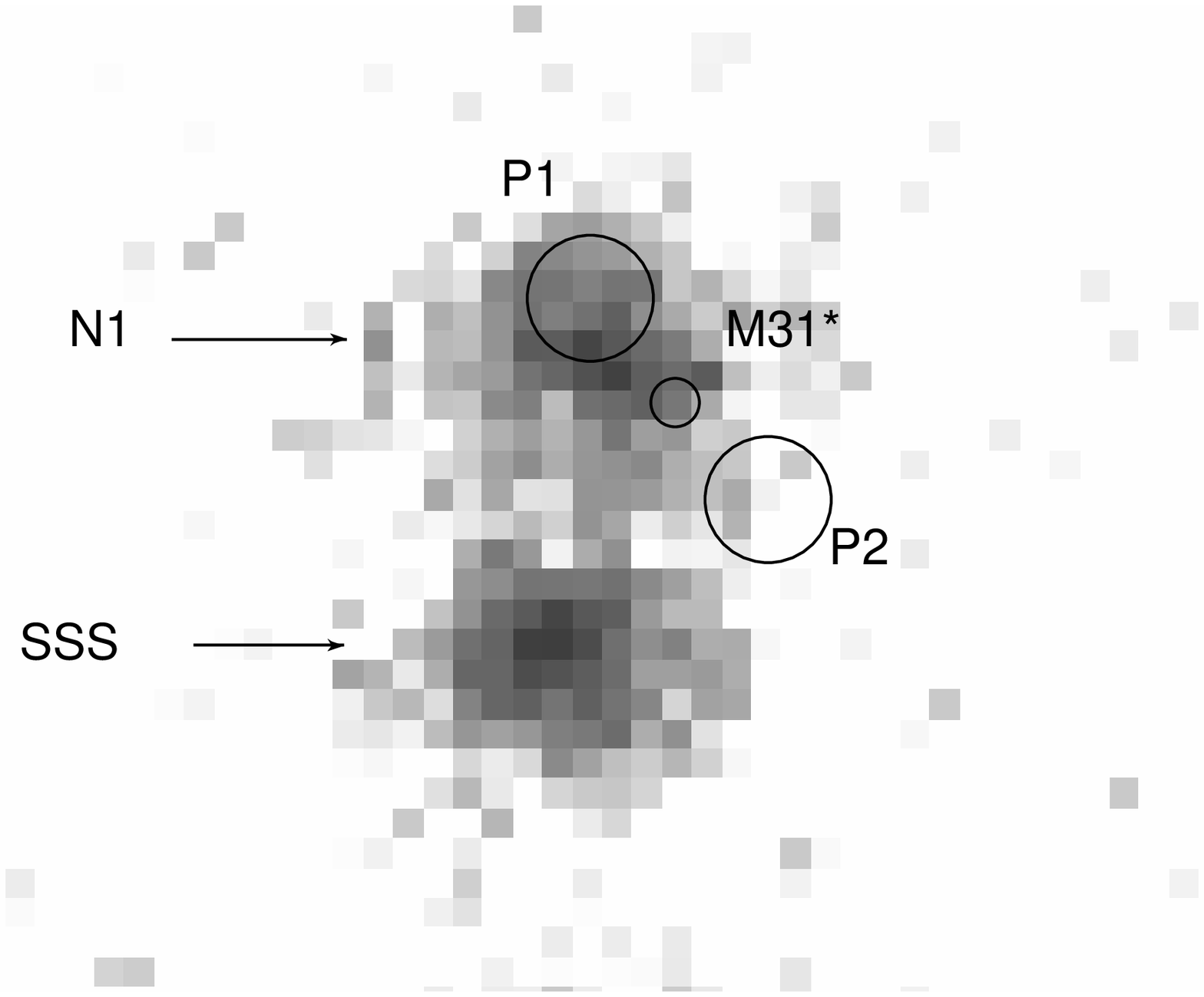}
\caption{The 35.7~ks merged ACIS image of the M31 nucleus with $1/8''$ pixels.
The sources labeled N1 and SSS are known as \none/r1-10 and \sss/r1-9 
in \citet{kong.2002}.   The distance between these two sources is
$1.2''$.  The supersoft source SSS is 
fainter in this epoch than in the HRC image. The position 
of M31* is marked with a small ($0.1''$ radius = $1\sigma$ position
error) circle.  The approximate locations of the diffuse double
nucleus P1/P2 are marked with larger circles.\label{acis3c}}
\end{figure}

\begin{figure}
\plotone{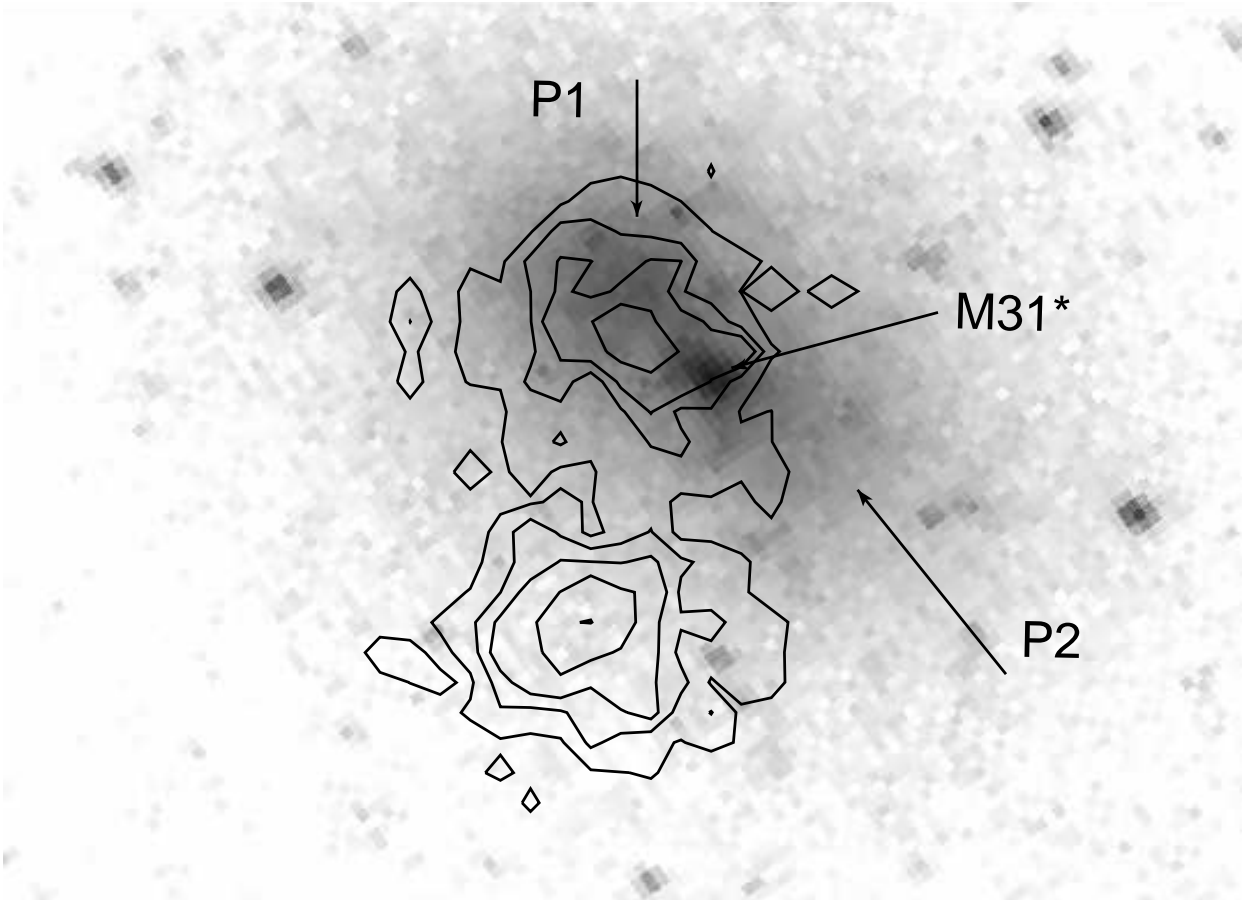}
\caption{
An archival F300W PC2 image of the
M31 nucleus with ACIS contours overlayed. The double nucleus P1/P2 is
marked, but P2 is only faintly discernible due
to its red color. 
M31* is the brightest object near the center. 
The ACIS contours show
that M31* is not co-incident with any of the bright X-ray sources.
While these contours do not show a resolved source at the position
of M31*, the slightly elongated contours in its direction are
consistent with the flux of M31* indicated by the HRC image.
\label{wfpc2.aciscont}
}
\end{figure}

\begin{figure}
\plotone{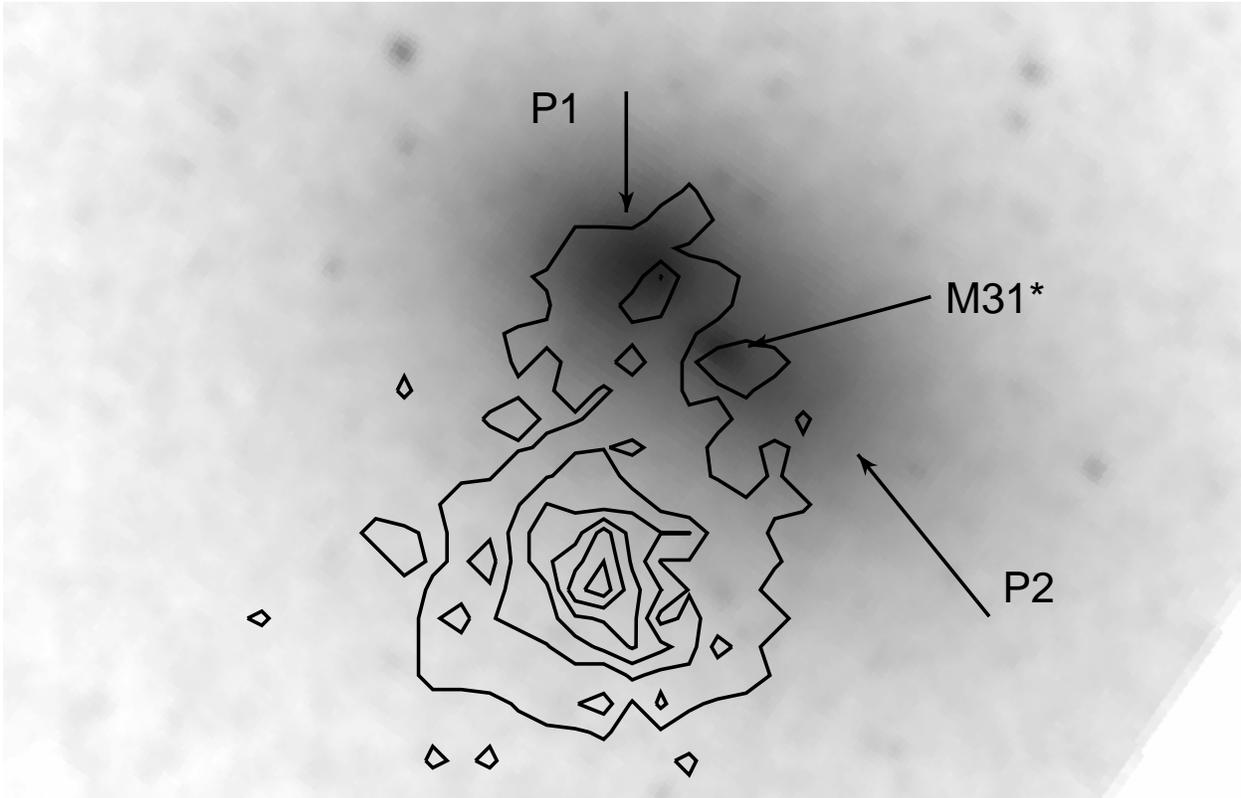}
\caption{The HST/ACS image of the M31 nucleus with HRC contours overlayed.  The
double nucleus P1/P2 is marked, but P2 is only faintly discernible on
this B-band image due to its red color.  
M31* is the more compact, bright object to the lower right of P1. 
There is a clear separate
contour consistent with the position of M31*, indicating a possible
X-ray detection of this SMBH.  The plate scale here is the same as in
Figure~2. 
\label{acs.hrccont}
}
\end{figure}

\begin{figure}\label{hrc.image}
\plotone{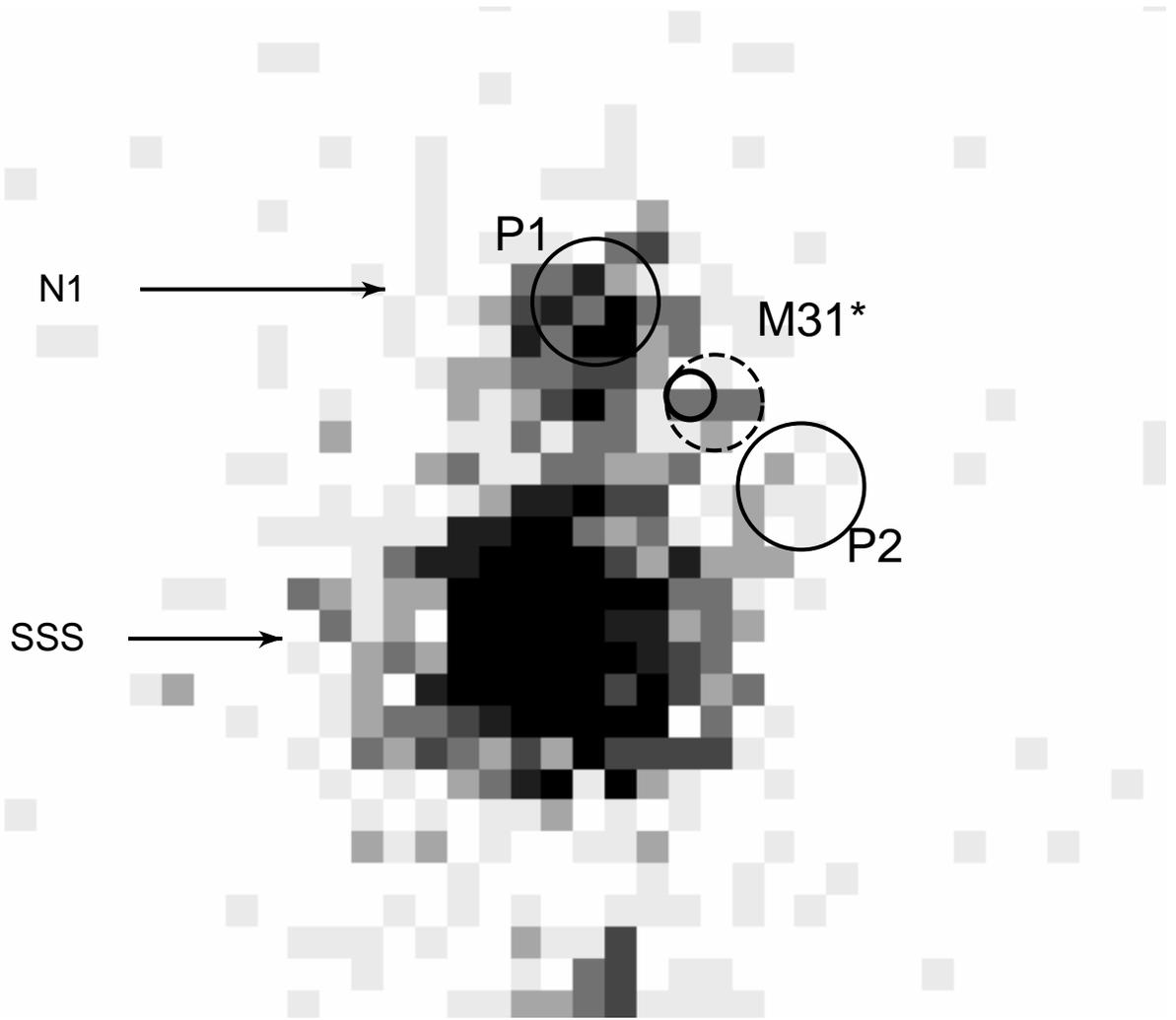}
\caption{
The 47~ks HRC image of the M31 nucleus, with $1/8''$ pixels. The source
at the top is N1, the brighter source in the center is the super-soft
source SSS.  The
position of M31* is marked with a small ($0.1''$ radius = 1$\sigma$
position error) heavy circle
in the center.  The source outlined with the dashed circle is clearly
consistent with the position of M31*.
There are 13 counts within this dashed 
circle. The locations of the diffuse double
nucleus P1/P2 are schematically indicated with the larger circles.  
}
\end{figure}

\begin{figure}\label{bondi.fig}
\plotone{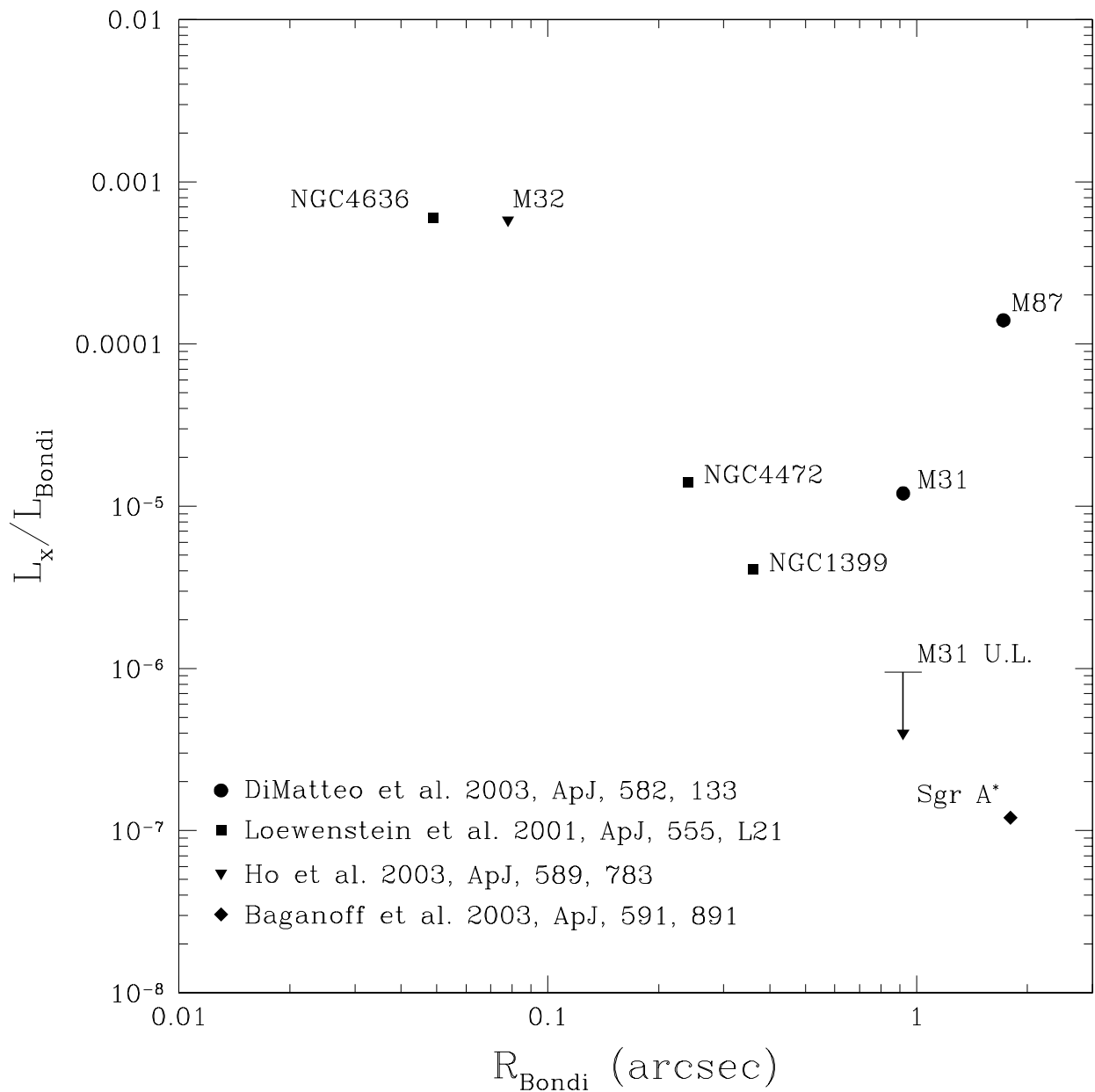}
\caption{
The Bondi radii of nearby SMBH vs. their apparent X-ray luminosity (or
upper limits) in units of the expected Bondi luminosity.  
The only two extra-galactic SMBH with resolvable radii are M31* and
M87. The point labeled M31 corresponds the the possible detection
reported herein, and the line labeled 'M31 U.L.' corresponds the upper
limit possible with a 200~ks \chandra\/ HRC-I observation. 
}
\end{figure}

\end{document}